\RequirePackage{amsmath,amssymb}
\documentclass{ifacconf}

\usepackage{graphicx}      
\usepackage{natbib}        

\usepackage{marvosym}
\usepackage{color}
\graphicspath{{Figures/}}
\usepackage{tikz}
\usetikzlibrary{arrows,automata}
\usepackage{bm}			  
\newcommand{\infnorm}[1]{\Vert #1 \Vert_{\infty}}

\usepackage[inline]{enumitem}
\usepackage{balance}
\begin{document}
\begin{frontmatter}
\title{
Approximate Abstractions of Markov Chains with Interval Decision 
Processes\thanksref{footnoteinfo}
(Extended Version)
} 
\thanks[footnoteinfo]{This research has been partially supported 
by the Alan Turing Institute, London, UK, by the ECSEL SafeCOP project n.692529, 
and by a grant from the Filauro Foundation.}
\author[First]{Yuriy {Zacchia Lun}}
\author[Second]{Jack Wheatley} 
\author[First]{Alessandro {D'Innocenzo}}
\author[Second]{Alessandro Abate}
\address[First]{Department of Information Engineering, Computer Science and 
	Mathematics, University of L'Aquila, and Center of Excellence DEWS, Italy 
	(e-mail: \{yuriy.zacchialun,alessandro.dinnocenzo\}@univaq.it)}
\address[Second]{Department of Computer Science, University of Oxford, UK \\ 
	(e-mail: alessandro.abate@cs.ox.ac.uk)}
\begin{abstract}                
This work introduces a new abstraction technique for reducing the state space of 
large, discrete-time labelled Markov chains. The abstraction leverages the 
semantics of interval Markov decision processes and the existing notion of 
approximate probabilistic bisimulation.
Whilst standard abstractions make 
use of abstract points that are taken from the state space of the concrete model 
and which serve as representatives for sets of concrete states, in this work the 
abstract structure is constructed considering abstract points that are not 
necessarily selected from the states of the concrete model, rather they are a 
function of these states. The resulting model presents a smaller one-step 
bisimulation error, when compared to a like-sized, standard Markov chain 
abstraction. We outline a method to perform probabilistic model checking,  
and show that the computational complexity of the new method is comparable to 
that of standard abstractions based on approximate probabilistic bisimulations. 
\end{abstract}
\end{frontmatter}
\section{Introduction}\label{sec:intro}
This work investigates new notions of probabilistic bisimulations of labelled, 
discrete-time Markov chains (LMCs), see \cite{larsen}, \cite{BP94}. 
It specifically focuses on new, forward approximate notions. There is a practical 
need to develop better approximations that are easily computable and utilisable 
in quantitative model checking procedures (see \cite{AdaptiveAggregation,SD06}): 
the principal objective of this work is indeed to develop tighter notions of 
approximate probabilistic bisimulations (APBs).  

A standard approach to improve the one-step error, which quantifies the difference 
between concrete and abstract transition probabilities, is to search an optimal 
lumping of the states of the concrete model (as a partition or a covering of its 
state space). Alternatively, whenever one is bound to work with a specific 
partitioning of the state space as dictated by the labelling of the states,  
one can select specific representative points of each class of
states that best represents the partition, namely that yields the smallest 
one-step error. 

In this work we develop a new approach to create abstractions of LMCs, given a 
fixed, label-preserving partition of its state space. Label-preserving partitioning 
is relevant for the study of properties expressed as temporal specifications defined 
over the labels of interest \cite{desharnais2002bisimulation}. Our approach
leverages the semantics of interval-valued labelled Markov chains (IMCs) and
Markov decision processes (MDPs). It relies on creating a set 
of representatives of each partition that 
are not necessarily states of the concrete model. Specifically, 
we consider all the states of the concrete model belonging to each partition and 
derive ``virtual'' states of the abstract model, which are function of the concrete 
states and which can achieve optimal APB error for a given probabilistic 
computational tree logic (PCTL) formula: such virtual states are obtained by 
considering the best possible transition probability vector within a related 
transition probability interval built from the concrete model. In doing so, we 
show that we can produce an abstract model that has a smaller (or equal, in the 
worst case) error compared to any possible abstraction obtained via the standard 
method from literature.  
We also show, as expected, that the bounds on the propagation of this error in time 
outperform similarly derived error propagation bounds for standard APB-based 
abstractions. We argue that our new approach is comparable to the standard abstraction 
algorithms in terms of the computation time required to perform probabilistic model 
checking over the abstract model. The derivation of an abstraction error over a time 
horizon, which accommodates for general PCTL formulae, allows to refine the outcomes 
of the model checking procedure performed with the abstract model, over the original 
concrete Markov chain. 

\noindent \textbf{Related Work}. 
This work taps into literature on approximate probabilistic bisimulations for 
robust PCTL model checking, see e.g.
\cite{RobustPCTLModelChecking}, \cite{puggelli2013polynomial}, and 
\cite{bian2017relationship}.
Our approximation notions are distinguished from related ones on approximation 
metrics over infinite traces, as in \cite{CK14}, \cite{TB16}. 
Further, we do not consider here 
Markov models with denumerable (cf. \cite{probability}) or uncountably 
infinite state spaces, as in \cite{bcAKNP14}, \cite{desharnais2004metrics}.
\pagestyle{headings}
\section{Preliminaries}
\noindent \textbf{Labelled Markov Chains.} The LMCs considered in this work are
discrete-time Markov chains with decorated states, as in \cite{katoen2008principles}, 
and are a subclass of labelled Markov processes as in \cite{desharnais2004metrics}. 
\begin{defn}
An \textit{LMC} is a tuple 
$\mathcal{D} \!=\! (S, P,L)$, where:
\begin{itemize}
\item $S$ is a non-empty, finite set of states;
\item $P: S \times S \rightarrow [0,1]$ is a transition probability matrix,
\item $L: S \rightarrow \mathcal{O}$ is a labelling function, where 
$\mathcal{O}\!=\!2^{\mathrm{AP}}$, and $\mathrm{AP}$ is a fixed set of atomic 
propositions, or labels. 
\end{itemize}
\end{defn}
For any state $s\!\in\!S$ and partition $Q\!\subseteq\!S$, we have that 
$P(s,Q)\!=\!\sum_{q_i \in Q}P(s,q_i)$. The function $L(s)$ captures all the 
observable information at state $s \!\in\! S$: this drives the bisimulation 
notion relating pairs of states, and we characterise properties over the 
codomain of this function. For a finite set of states $S$, $\Delta(S)$ 
denotes the set of distributions over $S$, i.e., the set of functions 
$p\!:\!S\!\rightarrow\![0,1]$ such that $\sum_{s\in S}p(s)\!=\!1$. 
Equivalently, we can think of $\Delta (S)$ as the set of stochastic vectors 
defined over $\mathbb{R}^{\vert S\vert}$.  
\begin{defn}
A \textit{path} in an LMC $\mathcal{D} \!=\! (S,P,L)$ is sequence of states 
$\omega \!=\! s_0,s_1,\dots, s_k$ where for all 
$i \!\in\! \mathbb{N} \cup \{0\}$, $s_i \!\in\! S$ and $P(s_i,s_{i+1})\!>\!0$. 
We denote by $\omega (i)$ the $(i\!+\!1)$-th state on the path $\omega$, and 
$\forall k\!\in\!\mathbb{N}$ we denote as $\mathrm{Paths}(s_0,k)$ the set of all 
paths in $\mathcal{D}$ of length $k\!+\!1$ and such that $\omega (0) \!=\! s_0$. 
\end{defn}
Notably, we consider paths of length $k\!+\!1$, 
made up of one initial state and precisely $k$ transitions. Letting 
$k\!\to\!\infty$ we obtain the definition of infinite path. 
\begin{defn}
For a finite path $\omega$ in an LMC $\mathcal{D}$, we define a 
\textit{cylinder set} as the set of all possible continuations of $\omega$, 
i.e., $\mathrm{Cyl}(\omega)\!\triangleq\!\{ \omega ^{\prime}\!\in\! 
\mathrm{Paths}(s_0,\infty)~\vert~\omega\text{ is a prefix of }\omega^{\prime}\}$. 
\end{defn}
For a given LMC $\mathcal{D} \!=\! (S,P,L)$, the transition probabilities from 
the matrix $P$ can be used to determine the probability of specific finite paths 
unfolding from a given state $s_0$ as follows. For any finite path 
$\omega \!\in\! \mathrm{Paths}(s_0,k)$ in $\mathcal{D}$ we introduce the function 
$\tau : \mathrm{Paths}(s_0,k) \!\to\! [0,1]$, defined as 
\begin{equation*}
\tau(\omega) \!\triangleq\! \begin{cases}  1 
	&\text{\!\!\!\!\!\!\!\!\!\!\! if the length of } \omega\text{ is one} \\ 
P(s_0,s_1)\cdot \dotsc \cdot P(s_{k-1},s_k)  &\text{otherwise.} \end{cases}
\end{equation*}
We hence define a probability space over all paths of the LMC $\mathcal{D}$ 
beginning from a state $s_0$, as follows:
\begin{itemize}
\item The sample space is $\Omega \!=\! \mathrm{Paths}(s_0, \infty)$.
\item The collection of events $\Sigma$ is the least 
$\sigma$-algebra on $\Omega$ containing $Cyl(\omega)$ for all finite paths 
$\omega$ starting at $s_0$.
\item The probability measure $\textbf{Pr}_{s_0}\!:\!\Sigma\!\rightarrow\![0,1]$ is 
defined by $\textbf{Pr}_{s_0}(Cyl(\omega)) \!=\! \tau(\omega)$ 
for any finite path $\omega$ beginning from a state $s_0$, which uniquely extends 
to a probability measure on the whole event space. 
\end{itemize}
For further details on probability spaces over paths in finite-state Markov chains, 
see \cite{probability}. 
\begin{defn}
Let $\mathcal{D}$ be a given LMC, with $s_0\!\in\!S$, and 
$k\!\in\!\mathbb{N}$. To any finite path $\omega$ 
in $\mathcal{D}$ corresponds a sequence of observations 
$\alpha(\omega) \!=\! L(\omega(0)), L(\omega(1)), \dots, L(\omega(k))$, which we 
call the \textit{trace} of $\omega$. Then, we denote by $\mathrm{Traces}(s_0,k)$ 
the set of all traces generated by the set $\mathrm{Paths}(s_0,k)$. 
\end{defn}
Let $\bm{\mathcal{T}}$ denote a set of traces of length $k\!+\!1$,
$\bm{\mathcal{T}}\!\subseteq\!\mathcal{O}^{k+1}$.
Any trace $\bar{\alpha}\!\in\!\bm{\mathcal{T}}$ is defined by a sequence of 
observations $\bar{\alpha}_0,\bar{\alpha}_1,\dots,\bar{\alpha}_k$, where an 
element $\bar{\alpha}_i$ is generated by a state $s_i\!\in\!S$ such that 
$L(s_i)\!=\!\bar{\alpha}_i$. Let $L^{-1}(\alpha_i)$ denote a partition of 
$S$ defined by the label $\bar{\alpha}_i$, 
$L^{-1}(\bar{\alpha}_i)\!\triangleq\!\{s\!\in\!S\,\vert\,L(s)\!=\!\bar{\alpha}_i\}$.
Thus, each trace $\bar{\alpha}$ is obtained from a corresponding set of paths
$\bm{\mathcal{P}}(\bar{\alpha})\!\triangleq\!\{\omega\!\in\!\mathrm{Paths}(s_0,k)
\,\vert\,\omega(i)\!\in\!L^{-1}(\bar{\alpha}_i)\}$, and we can 
quantify the probability that a given LMC $\mathcal{D}$ 
generates any of the traces in the set $\bm{\mathcal{T}}$, through a function 
$\varphi \!:\! \mathrm{Traces}(s_0,\!k) \!\!\to\!\! [0,\!1]$, 
as in \cite{bian2017relationship}):
\begin{defn}
$\!$Let $\mathcal{D}\!=\!(S,P,L)$ be a given LMC, with an initial distribution 
$p_0$ over $S$, $s_0\!\in\!S$, $k\!\in\!\mathbb{N}$. The probability that, 
starting from $s_0$, $\mathcal{D}$ generates any of the runs 
$\bar{\alpha}\!\in\!\bm{\mathcal{T}}\!\subseteq\!\mathrm{Traces}(s_0,k)$, is
$\varphi(s_0,\!\bm{\mathcal{T}})\!=\! p_0(s_0) 
\sum\nolimits_{\bar{\alpha} \in \bm{\mathcal{T}}} 
\sum\nolimits_{\bar{\omega} \in \bm{\mathcal{P}}(\bar{\alpha})}
\tau(\bar{\omega})$.
\end{defn}
We are interested in the verification of probabilistic properties of a given LMC, 
that can (and mostly will) be expressed in PCTL. The syntax and semantics of PCTL 
for LMCs are well known and will be just recalled next.

\noindent \textbf{Probabilistic Computational Tree Logic}.
Unlike the most standard definition of PCTL, see \cite{hansson1994logic}, 
we emphasise the role of the bounded-until operator, which is key in later parts 
of this work.
\begin{defn}
The \textit{syntax of PCTL} is as follows:
\begin{itemize}
\item $\phi ::= \mathrm{true} ~\vert ~ a ~\vert ~ \phi \wedge \phi ~\vert ~
		\neg \phi ~\vert ~ \mathbb{P}_{\sim p}[\psi]$, \hfill (state formulae)
\item $\psi ::= \textbf{X}\phi ~\vert ~ \phi \textbf{U}^{\,\leq k} \phi$, 
		\hfill (path formulae)
\end{itemize}
where $a$ is an atomic proposition, $\sim \, \in\! \{ <, >, \leq, \geq \}$, 
$p \!\in\! [0,1]$ is a given probability level, and 
$k \!\in\! \mathbb{N} \cup \{\infty\}$. A PCTL formula is defined to be a state formula.
\end{defn}
The semantics of PCTL over an LMC can be found in \cite{hansson1994logic}.
We have, for instance, that 
{$s\!\vDash\!\mathbb{P}_{\!\sim p}[\psi] \!\Leftrightarrow\!
\textbf{Prob}(s,\psi)\!\triangleq\!\textbf{Pr}_s\{\omega \!\in\! 
\textrm{Paths}(s,\infty) \,\vert\, \omega\!\vDash\!\psi\}\!\sim\! p$,}
where $s$ is a given state of the LMC.
The output of a model checking algorithm for a PCTL formula $\phi$ over LMC 
$\mathcal{D}$ is the set containing all the states of the model satisfying $\phi$ 
(see \cite{hansson1994logic}):
$\mathrm{Sat}(\phi) \!=\! \{ s \!\in\! S ~\vert~ s\vDash \phi \}.$

\noindent \textbf{Approximate Probabilistic Bisimulation}.
We recall the notion of APB of LMCs, and the related concept 
of approximate trace equivalence, as presented in \cite{bian2017relationship}. 
\begin{defn}
For a relation $\Gamma \!\subseteq\! S \!\times\! S$, we say that $Q \!\subseteq\! S$ 
is \textit{$\Gamma\!$-closed} if $\Gamma(Q) \!=\! 
\{ s \!\in\! S \,\vert\, \exists s^{\prime} \!\in\! Q \text{ such that } 
(s,s^{\prime})\!\in\!\Gamma\} \!\subseteq\! Q$.
\end{defn}
\begin{defn}\label{APBdef}
Given an LMC $\mathcal{D}$,
an APB with precision (error) $\varepsilon \!\in\![0,1]$ is a symmetric binary 
relation $\Gamma_{\varepsilon} \!\subseteq\! S \!\times\! S$ such that for any 
$(s, s^{\prime})\!\in\!\Gamma_{\varepsilon}$, 
one has $L(s)\!=\!L(s^{\prime})$, and for any 
$(s, s^{\prime})\!\in\!\Gamma_{\varepsilon}$ and $Q\!\subseteq\!S$, 
$P(s^{\prime},\Gamma_{\varepsilon}(Q))\!\geq\!P(s,Q)\!-\!\varepsilon$. 
Furthermore, two states $s,s^{\prime}\!\in\!S$ are said to be 
$\varepsilon$-bisimilar if there exists an APB $\Gamma_{\varepsilon}$ such that 
$(s, s^{\prime})\!\in\!\Gamma_{\varepsilon}$.
\end{defn}
We remark that an $\varepsilon$-APB is not an equivalence relation, since 
in general it does not satisfy the transitive property
(small approximation errors can accumulate to result in a large error). 
The last condition raised in Definition~\ref{APBdef} can be understood intuitively 
as ``for any move that $s$ can take (say, into set $Q\!\subseteq\!S$), $s^{\prime}$ 
can match it with higher likelihood over the corresponding set $\Gamma_{\varepsilon}(Q)$, 
up to $\varepsilon$ precision.''
\begin{defn}\label{def:trace_equivalence}
For a non-decreasing function $f \!:\! \mathbb{N} \!\to\! [0,1]$, we say that states 
$s, s^{\prime}$ of an LMC are $f(k)$-approximate probabilistic \textit{trace equivalent} 
if for all $k\!\in\!\mathbb{N}$ we have over 
$\bm{\mathcal{T}}\!\subseteq\!\mathcal{O}^{k+1}$ that 
$\vert\varphi(s,\!\bm{\mathcal{T}})\!-\!\varphi(s^{\prime}\!,\!\bm{\mathcal{T}})\vert\!\leq\!f(k).$
\end{defn}
\begin{thm}[\cite{bian2017relationship}]\label{theorem:bisimulation_implies_trace_equivalence}
If the states $s,s^{\prime}$ are $\varepsilon$-bisimilar, then
$s,s^{\prime}$ are $(1 \!-\! (1 \!-\! \varepsilon)^k)$-trace equivalent.
\end{thm}
\noindent \textbf{Standard Approach to LMC Abstractions.}
Given an LMC $\mathcal{D}=(S,P,L)$, consider a partition of the state space into 
subsets $S_1, \dots , S_m$, and such that for all $1 \!\leq\! i \!\leq m$, for all 
$s,s^{\prime} \!\in\! S_i$, $L(s)\!=\!L(s^{\prime})$. 
Let us denote the considered partitioning as $\bm{\mathcal{S}} \!=\! \{S_1,\dots , S_m\}$. 
Assume without loss of generality that the initial state is $s_0 \!\in\! S_1$. 
This partitioning induces a family of APBs as follows: 
\begin{enumerate}
\item Choose an element $s_i$ from each $S_i$: $\forall i\!\leq\!m$, $s_i \!\in\! S_i$. 
\item Define a relation $\Gamma_{\!\varepsilon}$ on $S \!\times\! S$: $\forall q\!\in\!S$,
$\Gamma_{\!\varepsilon} \!=\!\;\bigcup_{i=1}^{m}\{ (s_i,q),$ $(q,s_i) \,\vert\,$ $\!\!q \!\in\! S_i\}$.
This binary symmetric relation defines an APB, with an error determined by 
\begin{equation*}
\varepsilon = 
	\max\limits_{(s,s^{\prime}) \in \Gamma_{\varepsilon}}
	\max\limits_{\mathcal{Q} \in 2^{\bm{\mathcal{S}}}} 
		P(s, \mathcal{Q}) - P(s^{\prime},\Gamma_{\!\varepsilon}(\mathcal{Q})) .
\end{equation*}
Since $\forall i\!\leq\!m$, the chosen element $s_i \!\in\!S_i$, and 
$\Gamma_{\!\varepsilon}$ is a binary symmetric relation, we have by construction that
$\forall \mathcal{Q} \!\in\!2^{\bm{\mathcal{S}}}$,
$\Gamma_{\!\varepsilon}(\mathcal{Q})\!=\!\mathcal{Q}$. 
So, the sets $\{S_1,\dots,S_m\}$ are all non-intersecting, 
$\Gamma_{\!\varepsilon}$-closed and form a partition $\bm{\mathcal{S}}$ of $S$. 
Thus, we can write that
\begin{equation}\label{eq:vareps}
\varepsilon = 
	\max\limits_{(s,s^{\prime}) \in \Gamma_{\!\varepsilon}}
	\max\limits_{S_i \in \bm{\mathcal{S}}} 
		\vert P(s, S_i) - P(s^{\prime},S_i) \vert .
\end{equation}
\item From each of these APBs with errors $\varepsilon$ there is a corresponding 
lumped LMC $(\bm{\mathcal{S}}, P_{\Gamma_{\varepsilon}}, L)$ with an initial state 
$S_1$, where $\bm{\mathcal{S}} = \{S_1,\dots , S_m\}$; 
each $S_i$ is given the same label as its constituent elements in $\mathcal{D}$; 
and finally $P_{\Gamma_{\varepsilon}}$ is obtained as 
$P_{\Gamma_{\varepsilon}}(S_i,S_j) \!=\! P(s_i, S_j)$.
\end{enumerate}
A standard approach to reducing the error $\varepsilon$ of an APB for $\mathcal{D}$ 
is to choose the APB offering the lowest error within the family above. 
Notice that in this work we do not attempt to select alternative partitions and to 
optimise over them (see \cite{DLT08,APBalg}).  

We will compare and benchmark our new approach to generate abstractions of LMCs 
against this standard approach.   
The new approach will be introduced in Sect.~\ref{sec:abstractions_novel} and 
is based on the interval MDP (IMDP) semantics for 
IMCs, see \cite{IntervalMarkovChains}, 
as summarised next.

\noindent \textbf{Interval-Valued Labelled Markov Chains}.
\begin{defn}
Given non-negative matrices $A,B \!\in\! \mathbb{R}^{m \times n}$, with 
$A \!\leq\! B$ (elementwise), the transition set 
$\left[\Pi\right]\!=\!\left[A,B\right]$ is 
$\triangleq\!\big\{C \!\in\! \mathbb{R}^{m\times n} \,\vert\, C \text{ is stochastic, }
A \!\leq\! C \!\leq\! B \text{ (elementwise)\!}\big\}\!$.
\end{defn}
\begin{defn}
An IMC is a tuple $\mathcal{I}\!=\!(S,\left[\Pi\right]\!,L)$ where $S$ and $L$ 
are defined as for LMCs, and $\left[\Pi\right]\!=\!\left[P^l, P^u\right]$, with
$P^l, P^u \!:\! S \!\times\! S \rightarrow [0,1]$ matrices such that 
$P^l \!\leq\! P^u$ (element-wise) and  $P^l(s,s^{\prime})$ 
(respectively $P^u(s,s^{\prime})$) gives the lower (respectively upper) bound of 
the transition probability from state $s$ to $s^{\prime}$.
\end{defn}
In our novel LMC abstraction framework in Sect. \ref{sec:abstractions_novel},
we will be using transition sets defined by tight intervals.
\begin{defn}[\cite{hartfiel2006markov}]\label{def:tight}
Let $[u,v]$ be an interval of stochastic vectors $x$ defined by non-negative 
vectors $u$, $v$ in $\mathbb{R}^m$, with $u\!\leq\!x\leq\!v$ componentwise. 
For any component, indexed by $i$, if $u_i \!=\! \min\nolimits_{x \in [u,v]}x_i$ 
and $v_i \!=\! \max\nolimits_{x \in [u,v]}x_i$, then $u_i$ and $v_i$ are said to be tight. 
If all components $u_i$ and $v_i$ are tight, then we say that $[u,v]$ is tight.
\end{defn}
If an interval is not tight, one can always tighten it 
\cite[\textit{Tight Interval Algorithm}, p. 31]{hartfiel2006markov}:
given an interval $[u,v]$ of stochastic vectors in $\mathbb{R}^m$, 
the corresponding tight interval $[\bar{u},\bar{v}]$ of stochastic vectors is 
obtained by considering all components of the endpoints as follows:
\begin{itemize}[leftmargin=\parindent,align=left,labelwidth=\parindent,labelsep=0pt]
\item ~ if $u_i \!+\! \sum_{j \neq i}v_j \!\geq\! 1$, set $\bar{u}_i\!=\!u_i$; else, 
		set $\bar{u}_i\!=\! 1\!-\! \sum_{j \neq i}v_j$;
\item ~ if $v_i \!+\! \sum_{j \neq i}u_j \!\leq\! 1$, set $\bar{v}_i\!=\!v_i$; else, 
		set $\bar{v}_i \!=\! 1- \sum_{j\neq i}u_j$.
\end{itemize}
Noticeably, this algorithm produces a tight interval of stochastic vectors 
that contains exactly the same elements as the original one
\cite[Lemma 2.2, p. 31]{hartfiel2006markov}.
\begin{defn}[\cite{hartfiel2006markov}]\label{def:free}
Let $[u,v] \!\subset\! \mathbb{R}^m$ be a tight interval of stochastic vectors, 
and $x \!\in\! [u,v]$. We say that the \textit{element} $x_i$ of a stochastic 
vector $x$ is \textit{free}, if $u_i \!<\! x_i \!<\! v_i$.
\end{defn}
If $[u,v] \!\subset\! \mathbb{R}^m$ is a tight interval of stochastic vectors,   
then it defines a convex polytope over the Euclidean domain $\mathbb{R}^m$, and
a stochastic vector $x \!\in\! [u,v]$ is its vertex $\Leftrightarrow$ $x$ has at 
most one free element, say $x_i$, while all the other components $x_j$ of $x$,
$j\!\leq\!m$, $j\!\neq\!i$, coincide with the corresponding endpoints $u_j$ and 
$v_j$ (cf. \cite[Lemma 2.3, p. 32]{hartfiel2006markov}).
This result will be used in Sect.~\ref{sec:abstractions_novel}.

Next we recall the IMDP semantics of an IMC $\mathcal{I}$, where it is assumed 
that the transition probability matrix is chosen non-deterministically 
by the environment. Specifically, each time a state is visited, a transition 
distribution which respects the interval constraints is selected, and then a 
probabilistic step according to the chosen distribution is taken, see 
\cite{IntervalMarkovChains}:
\begin{defn}
The IMDP corresponding to an IMC $\mathcal{I}$ is the tuple 
$\tilde{\mathcal{I}} \!=\! (S, \delta, L)$, where $S$, $L$ are defined as for 
LMCs, and the set of actions $\delta \!:\! S \!\rightarrow\! 2^{\Delta (S)}$ 
is such that at any $s$ in $S$,
$\delta (s)\!=\!\{\mu \!\in\! \Delta(S) \,\vert\,\forall s^{\prime} \!\in\! S, 
P^l(s,s^{\prime}) \!\leq\! \mu (s^{\prime}) \!\leq\! P^u(s,s^{\prime})\}.$ 
\end{defn}
Even if, differently from classical MDPs, the action set $\delta (s)$ 
of an IMDP may contain infinitely many distributions, in \cite{IMCsSVA} it was 
shown that PCTL model checking for IMDPs can be reduced to model checking for MDPs:
\begin{thm}\label{imdp to mdp}
Given an IMDP $\tilde{\mathcal{I}}$, there exists an MDP $\mathcal{M}$ such that 
for any PCTL formula $\phi$, $\tilde{\mathcal{I}}\!\vDash\!\phi\!\Leftrightarrow\! 
\mathcal{M}\!\vDash\!\phi$. 
\end{thm}

So, given an IMC $\mathcal{I}$, a PCTL formula $\phi$, when considering the IMDP 
semantics for IMCs, we interpret $\tilde{\mathcal{I}} \!\vDash\! \phi$ in the 
same way as we interpret the relation $\vDash$ for classical MDPs. 

We recall that an MDP allows for probabilistic processes with 
non-deterministic choices (resolved by the environment or by an agent) characterised 
by the set of actions $\delta$ at each state.
In a standard MDP $\mathcal{M} \!=\! (S, \delta, L)$ the actions set 
$\delta(s)$ is finite and non-empty, and a path is defined to be a sequence of 
states and actions, $\omega\!=\!s_0,\mu_0,s_1,\mu_1,\dots,s_{k}$, where 
$\forall i\!\in\!\mathbb{N}$, $s_i\!\in\!S,\mu_i\!\in\!\delta(s_i)$ and 
$\mu_i(s_{i+1})\!>\!0$.
\begin{defn}
Given an MDP $\mathcal{M}$, a \textit{policy} (also known as strategy, 
or adversary) is a function $\sigma \!:\!\textrm{Paths}(s_0,k)\!\rightarrow\!\Delta(S)$
such that $\forall\omega\!=\!s_0,\mu_0,s_1,\mu_1,\dots,s_k$, 
$\sigma (\omega)\!\in\!\delta (s_k)$.
\end{defn}
A policy is memoryless if (and only if) the choice of action it makes at a state 
is always the same, regardless of which states have already been visited, as 
formally defined below 
\newpage
(adopted from \cite[p. 847]{katoen2008principles}):
\begin{defn}
A policy $\sigma$ on an MDP $\mathcal{M}$ is \textit{memoryless} (or simple)
$\Leftrightarrow$ $\forall\omega_1\!=\!s_0,\mu_0,s_1,\mu_1,\dots,
s_k\!\in\!\textrm{Paths}(s_0,k)$,
$\omega_2\!=\!q_0,\mu^{\prime}_0,q_1,\mu^{\prime}_1,\dots,
q_{\ell}\!\in\!\textrm{Paths}(q_0,\ell)$, with $s_k\!=\!q_{\ell}$, one has that
$\sigma(\omega_1) \!=\! \sigma(\omega_2)$.
\end{defn}
A memoryless policy of an MDP $\mathcal{M}$ can equivalently be defined as a 
state-to-action mapping at that state, i.e. a function
$\sigma\!:\!S\!\rightarrow\!\bigcup\nolimits_{s\in S}\delta(s)$,
where $\forall s\!\in\!S$, $\sigma(s)\!\in\!\delta(s)$.
It induces a finite-state LMC $\mathcal{M}^{\sigma} \!=\! (S,\tilde{P},L)$, with
$\tilde{P}$ being a stochastic matrix where the row corresponding to $s \!\in\! S$ 
is $\sigma(s)$.
Let us write $\textrm{Paths}^{\sigma}(s)$ for the infinite paths from $s$ where 
non-determinism has been resolved by an adversary $\sigma$, i.e. paths 
$s,\mu_0,s_1,\mu_1,\dots$ where 
$\forall k\!\in\!\mathbb{N}$, $\sigma(s,\mu_0,\dots,s_k)\!=\!\mu_k$.
So, for an MDP $\mathcal{M}$, a state $s \!\in\! S$, a path formula $\psi$ and a 
policy $\sigma$, we write 
$\mathbf{Prob}^{\sigma}(s,\psi)\!\triangleq\!
\textbf{Pr}_s\{\omega\!\in\!\textrm{Paths}^{\sigma}(s)~\vert~\omega\vDash\psi\}$. 
We further write 
$p_{\min}(s,\psi)\!\triangleq\!\inf_{\sigma}\mathbf{Prob}^{\sigma}(s,\psi)$ and 
$p_{\max}(s,\psi)\!\triangleq\!\sup_{\sigma}\mathbf{Prob}^{\sigma}(s,\psi)$.
\begin{defn}
Given an MDP $\mathcal{M}$ and a state $s$ of $\mathcal{M}$, 
for all non-probabilistic state formulae and path formulae the 
\textit{PCTL semantics} are identical to those for LMCs; moreover 
if $\sim \in \{\geq , >\}$, then 
$s\vDash \mathbb{P}_{\sim p}[\psi]\Leftrightarrow p_{\min}(s,\psi)\sim p$, while 
if $\sim \in \{\leq , <\}$, then 
$s\vDash \mathbb{P}_{\sim p}[\psi]\Leftrightarrow p_{\max}(s,\psi)\sim p$.
\end{defn}
As with LMCs, we will write $\mathcal{M}\vDash \phi$ if $s_0\vDash \phi$, where 
$s_0$ is a given initial state (or set thereof). Note that $\mathcal{M}\vDash\phi$ 
$\Leftrightarrow$ \textbf{for all} adversaries $\sigma$ at $s_0$, the LMC 
$\mathcal{M}^{\sigma}\vDash \phi$, and the output of a model checking algorithm 
for a PCTL formula $\phi$ over MDP $\mathcal{M}$ is the set containing all the 
states of the model satisfying $\phi$, i.e., 
$\mathrm{Sat}(\phi)\!=\!\{s\!\in\!S~\vert~s\vDash\phi\}$. 
\begin{thm}[\cite{katoen2008principles}]\label{memoryless}
Given an MDP $\mathcal{M}$, any state $s \!\in\! S$ and a PCTL path formula $\psi$, 
there exist memoryless adversaries $\sigma_{\min}$ and $\sigma_{\max}$ such that 
$\mathbf{Prob}^{\sigma_{\min}}(s,\psi) \!=\! p_{\min}(s,\psi)$,
$\mathbf{Prob}^{\sigma_{\max}}(s,\psi) \!=\! p_{\max}(s,\psi)$.
\end{thm}
In addition to verifying whether a PCTL state formula holds at a state in the model, 
we can also query the probability of the model satisfying a path formula, i.e. all 
of the values $\mathbf{Prob}(s,\psi)$, $p_{\min}(s,\psi)$, and $p_{\max}(s,\psi)$ 
are calculable via various existing model checking algorithms. For LMCs, we write 
this query as $\mathbb{P}_{=?}[\psi]$ for a given path formula, $\psi$, and for MDPs 
we have $\mathbb{P}_{\min=?} [\psi] \text{ and } \mathbb{P}_{\max=?} [\psi].$

\section{A New Abstraction Framework}\label{sec:abstractions_novel}
This section introduces the main contribution of this work, i.e. a novel 
approach to abstract LMCs, with the goal of obtaining an optimal precision in the 
introduced APB.   

\noindent \textbf{Transition Probability Rows with Optimal Error}.\\
We consider again an LMC $\mathcal{D}\!=\!(S,P,L)$ with partitioning 
$\bm{\mathcal{S}} \!=\!\{S_1, \dots , S_m\}$ of the state space $S$. 
We focus on the set $S_i \!=\! \{s_1^i,\dots, s_{n_i}^i\}$:
for $i\!\leq\!m$ and $\forall s_{\!j}^i\!\in\!S_i$, $j\!\leq\!n_i$, there is a
\begin{equation}\label{eq:vector_r}
\text{stochastic vector}\quad 
	r_{\!j}^{i} \!\triangleq\! (P(s_{\!j}^i, S_1),\dots,P(s_{\!j}^i, S_{m}))
\end{equation}

\vspace*{-2mm}
corresponding to the selection of element $s_{\!j}^i$ as abstraction point for 
the partition $S_i$. In the standard approach to constructing LMC abstractions 
described in Sect.~\ref{sec:intro}, 
to reduce the error of the abstraction one could choose $s_{\!j}^i\!\in\!S_i$ 
such that $\forall \ell \!\leq\!n_i$, 
$\varepsilon_{\!j}^{i}\!\triangleq\!
 \max\nolimits_{\ell\neq j} \infnorm{r_{\!j}^{i}\!-\!r_{\!\ell}^{i}}$ 
is minimised in $j$.\\
Thus, we can characterise the optimal error in terms of 
$\varepsilon_{\!j}^{i}$.
\begin{lem}\label{l1}
Given an LMC $\mathcal{D}$ and partitioning $\bm{\mathcal{S}}$, for any 
$s_{\!j}^i \!\in\! S_i$, let $r_{\!j}^{i}$ be defined by \eqref{eq:vector_r}.
Then the optimal (minimal) error achievable by $r_{\!j}^i$ is 
$\beta_i\!=\!\tfrac{1}{2} \varepsilon_{\max}^{i} \!=\! 
 \tfrac{1}{2} \max\limits_{1 \leq j \leq n_i} \varepsilon_{\!j}^{i}$, i.e.,
\begin{equation}\label{eq:epsilon_max}
\beta_i\!=\!\tfrac{1}{2}\max\limits_{1\leq j\leq n_i}\,\max\limits_{1\leq\ell 
\leq n_i,\, \ell \neq j} \infnorm{r_{\!j}^i \!-\! r_{\!\ell}^i}\!=\!
\varepsilon_{\min}^{i}.
\end{equation}
\end{lem}
\begin{pf}
See Appendix \ref{appendix:lemma1}.
\end{pf}
We can represent the set of stochastic vectors that achieve the optimal error value 
by a transition set, as follows (the proof is direct, by construction, and not reported here).
We denote by $\left(r_{\!j}^{i}\right)_{\!l}$ the $l$-th element of vector 
$r_{\!j}^{i}$, with $l\!\leq\!m$. 
\begin{prop}\label{prop1}
For $r_1^i, \dots , r_{n_i}^i$ as in \eqref{eq:vector_r}, and $\forall l\!\leq\!m$, let 
\begin{equation}\label{eq:interval_endpoints}
u_l^i \!\triangleq\!\! \min\limits_{1 \leq j \leq n_i} \left(r_{\!j}^i\right)_{\!l}, ~
v_l^i \!\triangleq\!\! \max\limits_{1 \leq j \leq n_i} \left(r_{\!j}^i\right)_{\!l}.
\end{equation}

\vspace*{-2mm}
Consider the transition set 
\begin{equation}\label{eq:inteval_def}
[u^i,v^i] \!\triangleq\! ([u_1^i,v_1^i], \dots , [u_{m}^i,v_{m}^i]).
\end{equation}

\vspace*{-2mm}
Let $\beta_i$ be defined by \eqref{eq:epsilon_max}.
Then the \textit{family of stochastic vectors} $r^i$ such that 
$\max\nolimits_{1 \leq j \leq n_i} \infnorm{r^i - r_j^i} \!=\! \beta_i$ 
is exactly the transition set defined as
\begin{equation}\label{eq:optimal_transition_set}
[u^i,v^i]_{\mathrm{opt}} \triangleq 
([v_1^i\!-\!\beta_i,u_1^i\!+\!\beta_i],\dots,[v_{m}^i\!-\!\beta_i,u_{m}^i\!+\!\beta_i]).
\end{equation}
\end{prop}
\begin{rem}\label{r3}
It is not always the case that $[u^i,v^i]_{\mathrm{opt}}$ is non-empty. 
As an example, consider rows
$r_{\!1}^i \!=\! (0.5, 0.3,0.2),$ $r_{\!2}^i \!=\! (0.45,0.33,0.22),$ and 
$r_{\!3}^i \!=\! (0.44,0.3,0.26)$. These rows are 3 linearly independent vectors 
in $\mathbb{R}^3$, with corresponding errors $\varepsilon_{\!1}^i\!=\!0.06$, 
$\varepsilon_{\!2}^i\!=\!0.05$, $\varepsilon_{\!3}^i\!=\!0.06$. If we consider 
in addition to these also the row $r_{\!4}^i \!=\! (0.45,0.34,0.21)$, we have that 
$\varepsilon_{\max}^i$ is still equal to $\infnorm{r_{\!1}^i\!-\!r_{\!3}^i}\!=\!0.06$, 
so the optimal error $\beta_i$ is equal to $0.03$. In this case, we have that 
$[u^i,v^i] \!=\! ([0.44,0.5],[0.3,0.34],[0.2,0.26])$ and hence 
$[u^i,v^i]_{\mathrm{opt}} \!=\! (0.47, [0.31,0.33], 0.23) \!=\! \emptyset $.
\end{rem}
A question naturally arises from the result of Remark~\ref{r3} on whether or not 
it is possible to reduce the error in cases where no stochastic vectors with 
associated error equal to $\varepsilon_{\min}^i$ exist. 
The response is affirmative, as shown next. 
\begin{prop}\label{prop2}
If for a set of $\mathbb{R}^{m}$-valued stochastic 
vectors $\{r_{\!1}^i,\dots, r_{n_i}^i\}$, each one defined by \eqref{eq:vector_r},
the corresponding set $[u^i,v^i]_{\mathrm{opt}}$, 
obtained through \eqref{eq:optimal_transition_set},
is empty, then there is a vector $r_{\ast}^i$ such that its corresponding error is
\begin{equation}\label{eq:gamma_i}
\varepsilon_{\ast}^i \!\leq\! 
\max \left\{\frac{1-\sum_{l=1}^{m} u_l^i}{m},\frac{\sum_{l=1}^{m} v_l^i-1}{m}\right\} 
\!=\!\gamma_i^{\prime}\!+\!\beta_i\!=\!\gamma_i.
\end{equation}
\end{prop}
\begin{pf}
See Appendix \ref{appendix:prop2}.
\end{pf}
\begin{exmp}\label{e6}
Consider again the rows 
$r_{\!1}^i$, 
$r_{\!2}^i$, 
$r_{\!3}^i$, 
$r_{\!4}^i$ from Remark~\ref{r3} that gave an empty $[u^i,v^i]_{\mathrm{opt}}$.
We have that $\sum_{l=1}^3 (v_l^i\!-\!\beta_i)\!=\!1.01$, 
$\gamma_i^{\prime}\!=\!\frac{1.01-1}{3}$, 
$r_{\ast}^i \!=\! \left(u_l^i\!-\!\gamma_i^{\prime}\right)_{l=1}^{m=3}$,
so $\varepsilon_{\ast}^i \!=\! 0.033\dot{3}$, which is only slightly worse than 
$\varepsilon_{\min}^i\!=\!\beta_i \!=\! 0.03$.
\end{exmp}
\noindent \textbf{Generation of the IMDP Abstraction}. 
One benefit of using transition sets to represent the set of vectors that optimally 
abstract the transition probabilities of each partition $S_i$ is that we can easily 
extend this to a family of transition probability matrices for the entire set of 
partitions: so the overall partition can either induce an optimal abstraction 
(in terms of the error, as per Proposition~\ref{prop1}) or one that is close to 
optimal (as per Proposition~\ref{prop2}). Specifically, let $\mathcal{D}\!=\!(S,P,L)$ 
be a given LMC, with a given initial state $s_0$, and state-space partition 
$S_1,\dots,S_m$. We obtain a procedure, to generate an IMC 
$[\mathcal{D}]\!=\!(\bm{\mathcal{S}},[\Pi],L)$, 
where 
$\bm{\mathcal{S}}=\{S_1,\dots,S_m\}$ is the lumped state space, and the 
initial 
condition is $S_1$ as described below. For all $1 \!\leq\! i \leq m$ we construct 
the transition set, $[\Pi]$ of the IMC as follows:
\begin{enumerate}
\item For $S_i = \{s^i_1, \dots ,s^i_{n_i}$\}, we obtain 
$r^i_1, \dots, r^i_{n_i}$ via \eqref{eq:vector_r}.
\item The minimal error achievable by any $r^i_{\!j}$ is $\beta_i$ from \eqref{eq:epsilon_max}. 
\item For all $l \!\leq\! m$, endpoints $u^i_l$, $v^i_l$ are defined by \eqref{eq:interval_endpoints}.
\item The transition set $[u^i, v^i]$ is obtained from \eqref{eq:inteval_def}.
\item The transition set $[u^i,v^i]_{\mathrm{opt}}$ is computed via \eqref{eq:optimal_transition_set}.
\item If $[u^i,v^i]_{\mathrm{opt}}$ is empty, find error ${\gamma_i}$ via \eqref{eq:gamma_i}.
\item Let $[u^i,v^i]_{\gamma_i}\!=\!
	([v^i_1\!-\!\gamma_i,u^i_1\!+\!\gamma_i],\dots,[v^i_{m}\!-\!\gamma_i,u^i_{m}\!+\!\gamma_i])$.
\item Let $R_i\!=\!\begin{cases}
	[u^i,v^i]_{\mathrm{opt}} &\text{ if }[u^i,v^i]_{\mathrm{opt}}\text{ is non-empty,} \\ 
	[u^i,v^i]_{\gamma_i} &\text{ otherwise.} \end{cases}$
\end{enumerate}
Let $[\Pi]$ be a matrix of intervals, with rows $R_1,\dots,R_m$,
which by construction is non-empty, so
$[\mathcal{D}]$ is a well-defined IMC, having associated one-step error
$\xi\!\triangleq\!(\xi_1,\dots,\xi_m)$, 
\begin{equation}\label{eq:error}
\xi_i = \begin{cases} \beta_i &\text{ if } [u^i,v^i]_{\mathrm{opt}} \text{ is non-empty,} \\ 
				     \gamma_i &\text{ otherwise.} \end{cases}
\end{equation}
\vspace*{-2mm}
The above procedure ensures that given any LMC, one can construct its unique 
optimal IMDP abstraction. The set of possible distributions one chooses from at 
each state are then from the set of vectors with optimal error.
\begin{defn}
Given an LMC $\mathcal{D}$, an \textit{IMDP abstraction} (IMDPA) of $\mathcal{D}$ 
(with associated error $\xi$) is the unique IMDP constructed using the 
IMC construction procedure.
\end{defn}
Recalling from Theorem~\ref{imdp to mdp} that PCTL model checking for IMDPs can 
be reduced to PCTL model checking for MDPs, we now give the following definition, 
remarking first that two models are said to be PCTL-equivalent $\Leftrightarrow$
they verify the same PCTL formulae.  
\begin{defn}
Given an LMC $\mathcal{D}$, an \textit{MDP abstraction} (MDPA) of $\mathcal{D}$ 
(with associated error $\xi$) is an MDP that is PCTL-equivalent to the IMDPA 
corresponding to $\mathcal{D}$.
\end{defn}
For a state $S_i$ of the IMDPA with transition probabilities within the transition 
set $[u^i,v^i]$, the corresponding state of the MDPA has action set equal to the 
set of the vertices of a convex hull $\mathrm{conv}([u^i,v^i])$, and hence all 
the actions are still points with optimal error. 
We can then perform model checking over this MDPA, knowing that the error at each 
state is still optimal in relation to the concrete model. 
We are then interested in 
determining how the probabilities of PCTL 
path formulae holding over the new abstraction compare to those over the 
concrete model. When considering whether a PCTL state formula of 
the form $\mathbb{P}_{\sim p}[\psi]$ is verified by an MDP at its initial state 
$s_0$, we need to calculate the values $p_{\min}(s_0,\psi)$ or $p_{\max}(s_0,\psi)$. 
We have introduced non-determinism into the abstracted model by considering it as an 
MDP, but we take control over the choice of actions at each state and hence always 
ensure that we choose a policy that achieves either the maximum or minimum probabilities. 
\begin{rem}\label{adverse}
As we have control over the choice of actions, we can assume a slight 
variation in PCTL semantics for IMCs presented in \cite{IntervalMarkovChains}. 
Typically, given an IMC $\mathcal{I}$ and a PCTL formula $\phi$, 
under the IMDP semantics, for $\tilde{\mathcal{I}}\!=\!(S,\delta, L)$, 
$\tilde{\mathcal{I}}\!\vDash\!\phi$ $\Leftrightarrow$ \textbf{for all} adversaries, 
$\sigma$, at $s_0$, $\tilde{\mathcal{I}}^{\sigma}\vDash \phi$. 
Since we have choice over the policy and over the initial condition of the IMDPA, 
we can argue that $\tilde{\mathcal{I}}\vDash \phi$ $\Leftrightarrow$ 
\textbf{there exists a} policy $\sigma$, which at $s_0$ is such that 
$\tilde{\mathcal{I}}^{\sigma}\vDash \phi$.
\end{rem}
\begin{defn}\label{def:imdpa_semantics}
Let $\tilde{\mathcal{I}}\!=\!(S,\delta, L)$ be the \textit{IMDPA} of a given LMC, 
with one-step abstraction error $\xi$.
Consider a state $s$ of $\tilde{\mathcal{I}}$. 
The \textit{PCTL semantics} are 
\begin{itemize}
\item if $\sim \in \{\geq,>\}$, then 
	$s\vDash\mathbb{P}_{\sim p}[\psi]\Leftrightarrow p_{\max-\epsilon_k}(s,\psi)\sim p$,
\item if $\sim \in \{\leq,<\}$, then 
	$s\vDash\mathbb{P}_{\sim p}[\psi]\Leftrightarrow p_{\min+\epsilon_k}(s,\psi)\sim p$,  
\end{itemize}
where $\epsilon_k$ is the abstraction error $\xi$ propagated at the $k$-th time step 
(in any bounded probabilistic formula). 
\end{defn}
An example of the derivation of $\epsilon_k$ will be provided 
at the end of this section.
The presented semantics ensure that the satisfiability of a formula on IMDPA 
guarantees its satisfiability on the original LMC. 

\noindent \textbf{Geometric Interpretation of the New Abstractions}.
We have seen from the IMC construction procedure 
that 
$\forall S_i$ of the lumped LMC we have vectors $r^i_1, \dots, r^i_{n_i}$ and the 
corresponding row of intervals $R_i$. 
Here we are interested in looking at the shape of $R_i$ in relation to the convex 
polytope defined by these vectors, namely $\mathrm{conv}(\{r^i_1, \dots, r^i_{n_i} \})$.  
The vectors $r_1^i,\dots , r_{n_i}^i$ all lie in $\mathbb{R}^{m}$. In particular,
they are members of the set of $m$-dimensional stochastic vectors, which is an 
$(m\!-\!1)$-dimensional simplex. We denote it by $\textbf{1}^{m}$. In $\mathbb{R}^2$ 
it is a line segment, in $\mathbb{R}^3$ a bounded plane, and so on \cite[pp. 32--33]{boyd2004convex}. 
Consider the interval $[u^i,v^i]_{\mathrm{opt}}$ of $R_i$, and let 
$\tilde{R_i}\!\triangleq\!\{t\!\in\!\mathbb{R}^{m}\,\vert\,\forall 1\!\leq\!j\!\leq\!m,\,v_{\!j}^i\!-\!\beta_i
\!\leq\!t_j\!\leq\!u_{\!j}^i\!-\!\beta_i\}$.
$\tilde{R}_i$ is an 
${m}$-dimensional hypercube, and 
$R_i \!=\! \tilde{R_i} \!\cap\! \textbf{1}^{m}$.
We can hence gain some intuition about $R_i$ in terms of this relationship. 
We know from Remark~\ref{r3} that 
$R_i \!=\![u^i,v^i]_{\mathrm{opt}}$ can be empty. 
This is not the case when
$\sum_{k=1}^{m} (v^i_k \!-\! \beta_i) \!\leq\! 1 \!\leq\! \sum^{m}_{k=1} (u^i_k \!+\! \beta_i)$, i.e., when 
$\sum_{k=1}^{m} v^i_k \!-\! m \beta_i \!\leq\! 1 \!\leq\! \sum_{k=1}^{m} u^i_k \!+\! m \beta_i$.
Intuitively, this happens if some point of the hypercube lies on one side of the simplex, and others are on the other side, 
thus yielding a non-empty intersection $R_i$.
Since the intersection of a family of convex sets is convex \cite[p. 81]{boyd2004convex}, when 
$R_i$ is non-empty, it is a 
convex polytope in $\mathbb{R}^{m}$. Thus we know that it must either have cardinality 1, or be uncountable infinite. 
From this geometric viewpoint, we can immediately identify the case where $\vert R_i \vert \!=\! 1$.
It happens when either a single corner of the hypercube touches the simplex, 
or when the hypercube is in fact a line that intersects the simplex.
We formally characterise the cases where $\vert R_i \vert \!=\! 1$
in the following lemma. 
\begin{lem}\label{size1}
Let $T=([t_1,w_1] , \dots , [t_{m}, w_{m}])$ be any transition set. Then 
$\vert T \vert \!=\! 1$ $\Leftrightarrow$ at least one of the following 
conditions holds:
\begin{enumerate*}
\item $\sum_{k=1}^{m} t_k \!=\! 1$, ~~
\item $\sum_{k=1}^{m} w_k \!=\! 1$, ~~
\item $\sum_{k=1}^{m} t_k \!<\! 1 \!<\! \sum_{k=1}^{m} w_k$, 
	$T$ has only one free element.
\end{enumerate*}
\end{lem}
\begin{pf}
See Appendix \ref{appendix:lemma2}.
\end{pf}
We can hence apply Lemma~\ref{size1} to 
$R_i \!=\! [u^i,v^i]_{\mathrm{opt}}$
to determine exactly the cases in which 
$\vert R_i \vert \!=\! 1$. Geometrically, 
we see that conditions (1) and (2) of Lemma~\ref{size1} are cases when the hypercube 
just touches the simplex with one of its vertices, and condition (3) is the case in which 
the hypercube is a line which intersects with the simplex. 

Beyond the case when $\vert R_i \vert \!=\! 1$, 
we have already seen that if $R_i$ is non-empty it must be uncountably infinite. 
This happens when the hypercube completely intersects (i.e. 
not in just a single point) the simplex. 
If we perform the Tight Interval Algorithm \cite[p. 31]{hartfiel2006markov} summarized in Sect.~\ref{sec:intro}, 
we obtain the smallest (in terms of volume) hypercube $T$ such that 
$T \cap \textbf{1}^{m} \!=\! \tilde{R_i} \cap \textbf{1}^{m}$.
In that case the vertices of the polytope $T \cap \textbf{1}^{m}$ will be a subset of the vertices of the hypercube $T$. 

Finally, when a hypercube has empty intersection with the 
simplex, that is,
$\tilde{R}_i \cap \textbf{1}^{m} \!=\! \emptyset$, we have that 
$\vert R_i \vert \!=\! 0$. 
In this case, 
we can relax the intervals defining $R_i$ by replacing $\beta_i$ with 
$\gamma_i$ defined by \eqref{eq:gamma_i}, 
which is enough to ensure non-emptiness. 
This process is equivalent to uniformly expanding the hypercube $\tilde{R}_i$ in all directions 
up until it touches the simplex 
$\textbf{1}^{m}$.
Formally, when $\gamma_i \!=\! \frac{\sum_{k=1}^{m} v_k^i-1}{m}$, we have that
$\sum_{k=1}^{m} (v_k^i \!-\! \gamma_i) \!=\!1$.
Hence, by Lemma~\ref{size1}, it has cardinality equal to 1.
The result is the same if we instead let $\gamma_i \!=\!\frac{1 - \sum_{k=1}^{m} u_k^i}{m}$.

\noindent \textbf{Complexity of Model Checking}.
It is well known that time complexity of model checking for LMCs is linear in size 
$\vert \phi \vert$ of a PCTL formula $\phi$ and polynomial in the size of the model. 
Specifically, for an LMC $\mathcal{D}$, denote by $\vert \mathcal{D} \vert$ the 
size of $\mathcal{D}$, i.e., the number of states, $m$, plus the number of pairs 
$(s,s')\!\in\!S\!\times\!S$ such that $P(s,s')\!>\!0$ \cite[p. 748]{katoen2008principles}. 
Then the PCTL model-checking problem $\mathcal{D}\!\vDash\!\phi$ can be solved in time
$\mathcal{O}(\mathrm{poly}(\vert\mathcal{D}\vert)\!\times\!\vert\phi\vert\!\times\!k_{max})$,
where $k_{max}$ is the maximal step bound that appears in a subpath formula 
$\phi_1\textbf{U}^{\leq k}\phi_2$ of $\phi$, and $k_{max}\!=\!1$ if $\phi$ does not 
contain a step-bounded until operator \cite[Theorem 10.40, p. 786]{katoen2008principles}.

We refer to \cite[pp. 530--531, 534]{puggelli2013polynomial} for the 
details on deriving the time complexity of PCTL model checking for the IMDPs 
$\tilde{\mathcal{I}}$, that is $\mathcal{O}(\mathrm{poly}(\vert
\tilde{\mathcal{I}}\vert)\!\times\!\vert\phi\vert\!\times\!k_{max})$,
where $k_{max}$ is defined as for LMCs, while the size $\vert\tilde{\mathcal{I}}\vert$ 
of an IMDP model for the purpose of model checking is $\mathcal{O}(m^2)$, 
where $m$ is the number of the states in the set $S$.

The PCTL model checking algorithm for IMDPA of an LMC is then based on routines 
presented in \cite{puggelli2013polynomial}: for non-probabilistic boolean operators, 
the verification step is standard, see \cite{katoen2008principles}, while for the 
probabilistic operators the satisfiability sets are computed by generating and then 
solving a number of convex optimisation problems.

\noindent \textbf{Error Propagation}.
At the beginning of this section, we have seen that by constructing an IMDPA 
of an LMC we can reduce the one-step error compared to any lumped LMC obtained 
via a standard approach to constructing abstractions based on APBs. In practice, 
the properties we wish to study are not just measurements of the one-step 
probabilities of the concrete model, but rather PCTL properties over a 
certain time horizon. 
Understanding how the error introduced by an abstraction propagates at increasing 
time steps is the key, so the focus here
is on studying how the error associated to IMDPAs evolves compared to that of the 
standard APB abstraction. 
Leveraging work done in \cite{bian2017relationship}, 
we find that the decreased one-step error leads to decreased error bounds across 
all finite (or infinite) time frames. 
We know from Sect.~\ref{sec:intro} that an APB with precision $\varepsilon$ 
(as in Definition~\ref{APBdef}) induces an upper bound on the probabilistic trace 
distance, quantifiable as $(1\!-\!(1\!-\!\varepsilon)^k)$, as per Theorem 
\ref{theorem:bisimulation_implies_trace_equivalence}. 

\newpage
This translates to a 
guarantee on all the properties implied by $\varepsilon$-trace equivalence, 
such as closeness of satisfaction probabilities over bounded-horizon linear time 
properties. 
Then, we know from Theorem~\ref{memoryless} that for any MDP there are memoryless 
policies for which the probabilities of the property holding are equal to the 
minimum and maximum probabilities across all possible policies, see 
\cite{katoen2008principles}. 
As these policies are memoryless, the LMC induced by them is finite state, and furthermore in the case of those obtained from our IMDPA, will be one-step $\xi$-error bisimulations of the concrete model. We hence just apply Theorem~\ref{theorem:bisimulation_implies_trace_equivalence} to these induced LMCs, and the result follows.
Thus, the abstraction error propagated at the $k$-th time step is $\epsilon_k \!\leq\! (1 \!-\! (1 \!-\! \xi)^k)$.

In the next section we will see, 
on a small case study,  
how the probabilistic safety properties expressed in PCTL may be checked on an IMDPA of an LMC,
and we will quantify the associated abstraction error.

\section{Case study}\label{sec:case_study}
In this section we compare the different LMC abstractions discussed above for model checking. 
We first illustrate how an abstracted model is found by using the standard approach from Sect.~\ref{sec:intro}, 
then show how IMDP and MDP abstractions are created by the new approach. 
We begin with the concrete model, an 11-state LMC $\mathcal{D} \!=\! (S,P,L)$, 
with states $s_0,\dots , s_{10}$, 
initial state $s_0$, 
and transition probabilities given by the following matrix $P$ equal to

\vspace*{-4mm}
\begin{small}
\begin{equation*}
\begin{pmatrix} 
0.05\, & 0.05\, & 0.05\, & 0.05\, & 0.15\, & 0.15\, & 0.15\, & 0.3\, & 0.02\, & 0.01\, & 0.02\, \\ 
0.04 & 0.04 & 0.05 & 0.05 & 0.14 & 0.17 & 0.15 & 0.28 & 0.03 & 0.03 & 0.02 \\
0.01 & 0.01 & 0.1 & 0.05 & 0.14 & 0.15 & 0.15 & 0.2 & 0.19 & 0 & 0 \\
0.06 & 0.04 & 0.06 & 0.07 & 0.16 & 0.17 & 0.15 & 0.07 & 0.03 & 0.05 & 0.14 \\
0.01 & 0.01 & 0 & 0 & 0.96 & 0 & 0 & 0.005 & 0.005 & 0.005 & 0.005 \\
0 & 0.01 & 0.01 & 0.01 & 0.01 & 0.95 & 0.01 & 0 & 0 & 0 & 0 \\
0 & 0 & 0 & 0 & 0.25 & 0.5 & 0.25 & 0 & 0 & 0 & 0 \\
0.15 & 0.15 & 0.15 & 0 & 0.15 & 0.15 & 0.15 & 0.02 & 0.03 & 0.03 & 0.02 \\
0.15 & 0.15 & 0.06 & 0.06 & 0.14 & 0.13 & 0.15 & 0.04 & 0.03 & 0.03 & 0.06 \\
0.4 & 0.04 & 0.02 & 0.01 & 0.14 & 0.13 & 0.15 & 0.1 & 0.01 & 0 & 0 \\
0.44 & 0 & 0 & 0 & 0 & 0 & 0.43 & 0 & 0 & 0 & 0.13 \\
\end{pmatrix}\!\!.
\end{equation*}
\end{small}

\vspace*{-3mm}
The labelling is the following: 
{\small $L(s_0) \!=\! L(s_1) \!=\! L(s_2) \!=\! L(s_3) \!=\! a$}, 
{\small $L(s_4) \!=\! L(s_5) \!=\! L(s_6) \!=\! b$}, and
{\small $L(s_7) \!=\! L(s_8) \!=\! L(s_9) \!=\! L(s_{10}) \!=\! c$}. 
We assume that the labelling induces the state-space partition, so we have that 
{\small $S_a \!=\! \{s_0, s_1, s_2, s_3\}$}, 
{\small $S_b \!=\! \{s_4, s_5, s_6\}$}, and 
{\small $S_c \!=\! \{s_7, s_8, s_9, s_{10}\}$}. 
From this lumping we obtain the following rows corresponding to each element of each set:

\begin{itemize}[leftmargin=\parindent,align=left,labelwidth=\parindent,labelsep=1pt]
\begin{small}
\item[$S_a$:] $(\!\!\;0.2\!\!\;,0.45\!\!\;,0.35\!\!\;)\!\!\;, 
			   (\!\!\;0.18\!\!\;, 0.46\!\!\;, 0.36\!\!\;)\!\!\;, 
			   (\!\!\;0.17\!\!\;,0.44\!\!\;, 0.39\!\!\;)\!\!\;, 
			   (\!\!\;0.23\!\!\;,0.48\!\!\;,0.29\!\!\;)\!\!\;.$
\item[$S_b$:] $(\!\!\;0.02\!\!\;, 0.96\!\!\;, 0.02\!\!\;)\!\!\;,
			   (\!\!\;0.03\!\!\;, 0.97\!\!\;, 0\!\!\;)\!\!\;, 
			   (\!\!\;0\!\!\;, 1\!\!\;, 0\!\!\;)\!\!\;.$
\item[$S_c$:] $(\!\!\;0.45\!\!\;, 0.45\!\!\;, 0.1\!\!\;)\!\!\;,
			   (\!\!\;0.42\!\!\;, 0.42\!\!\;, 0.16\!\!\;)\!\!\;, 
			   (\!\!\;0.47\!\!\;, 0.42\!\!\;, 0.11\!\!\;)\!\!\;, 
			   (\!\!\;0.44\!\!\;, 0.43\!\!\;, 0.13\!\!\;)\!\!\;.$
\end{small}
\end{itemize}

Following the standard approach to producing abstractions by choosing the rows 
from each set above with the best error, we obtain a 0.06-bisimulation of the 
model $\mathcal{D}$, where the representative points are $s_0\!\in\!S_a$, 
$s_5\!\in\!S_b$, and $s_{10}\!\in\!S_c$. The related transition probabilities are 
given by
\begin{small}
\begin{equation*}
P_{\Gamma_{\!\varepsilon}} \!=\!
\begin{pmatrix}
0.2 & 0.45 & 0.35\\
0.03 & 0.97 & 0\\
0.44 & 0.43 & 0.13
\end{pmatrix}\!.
\end{equation*}
\end{small}
Then, we apply the procedure from Sect.~\ref{sec:abstractions_novel}
to produce the IMDPA of $\mathcal{D}$, and get the optimal errors 
$\beta_a \!=\! 0.05$, $\beta_b \!=\! 0.02$, and $\beta_c \!=\! 0.03$ for the respective 
collections of rows $S_a$, $S_b$, and $S_c$. The related transition sets of abstracted
points with optimal error for each collection of rows are 
{$[u^a\!,v^a]_{\mathrm{opt}} \!=\! ([0.18,0.22], [0.43, 0.49], 0.34)$}, 
{$[u^b\!,v^b]_{\mathrm{opt}} \!=\! ([0.01,0.02],0.98, [0, 0.02])$}, and 
{$[u^c\!,v^c]_{\mathrm{opt}} \!=\! ([0.44,0.45],$ $[0.42, 0.45], 0.13)$},
respectively, which are all non-empty. 
We hence obtain an IMC $\mathcal{I} \!=\! (\{S_a,S_b,S_c\}, [\Pi], L)$, where $[\Pi]$ is the transition set represented by the $3\times 3$ matrix of intervals with rows $[u^a,v^a]_{\mathrm{opt}}$, $[u^b,v^b]_{\mathrm{opt}}$, $[u^c,v^c]_{\mathrm{opt}}$, and $L(S_a) \!=\! a$, $L(S_b) \!=\! b$, $L(S_c) \!=\! c$. We hence get that IMDPA is
$\tilde{\mathcal{I}} = (\{S_a,S_b,S_c\}, \delta, L)$,
where $\delta$ is defined for $i\!=\!a,b,c$ as: $\delta (S_i) \!=\! [u^i,v^i]_{\mathrm{opt}}$. The initial abstracted state is $S_a$. 

In order to perform the model-checking via standard tools for finite-state models 
such as PRISM (see \cite{kwiatkowska2011prism}), we construct the corresponding 
MDP abstraction {$\mathcal{M} \!=\! (\{S_a,S_b,S_c\}, \delta^{\prime}, L)$} 
from $\tilde{\mathcal{I}}$ by taking $\delta^{\prime}(S_i)$  
to be the set of vertices of the convex hull of $\delta(S_i)$ -- a procedure which 
can be found in \cite[p. 40]{hartfiel2006markov}. 

Next, we consider a bounded global property $\textbf{G}^{\leq k}\neg c$,
for values of $k$ ranging between $1$ and $20$. 
Here $\textbf{G}^{\leq k} \phi$ is a common shorthand for the path formula
$\neg (\mathrm{true} \textbf{U}^{\leq k}\neg \phi)$. 
Satisfaction of the PCTL formula $\mathbb{P}_{\sim p}[\textbf{G}^{\leq k}\phi]$ can be derivable as, 
for instance, 
$s\vDash \mathbb{P}_{\leq p}[\textbf{G}^{\leq k}\phi] \Leftrightarrow s\vDash \mathbb{P}_{\geq 1-p}[\mathrm{true} \textbf{U}^{\leq k}\neg\phi]$. 

\begin{figure}
\centering
\includegraphics[width=\columnwidth]{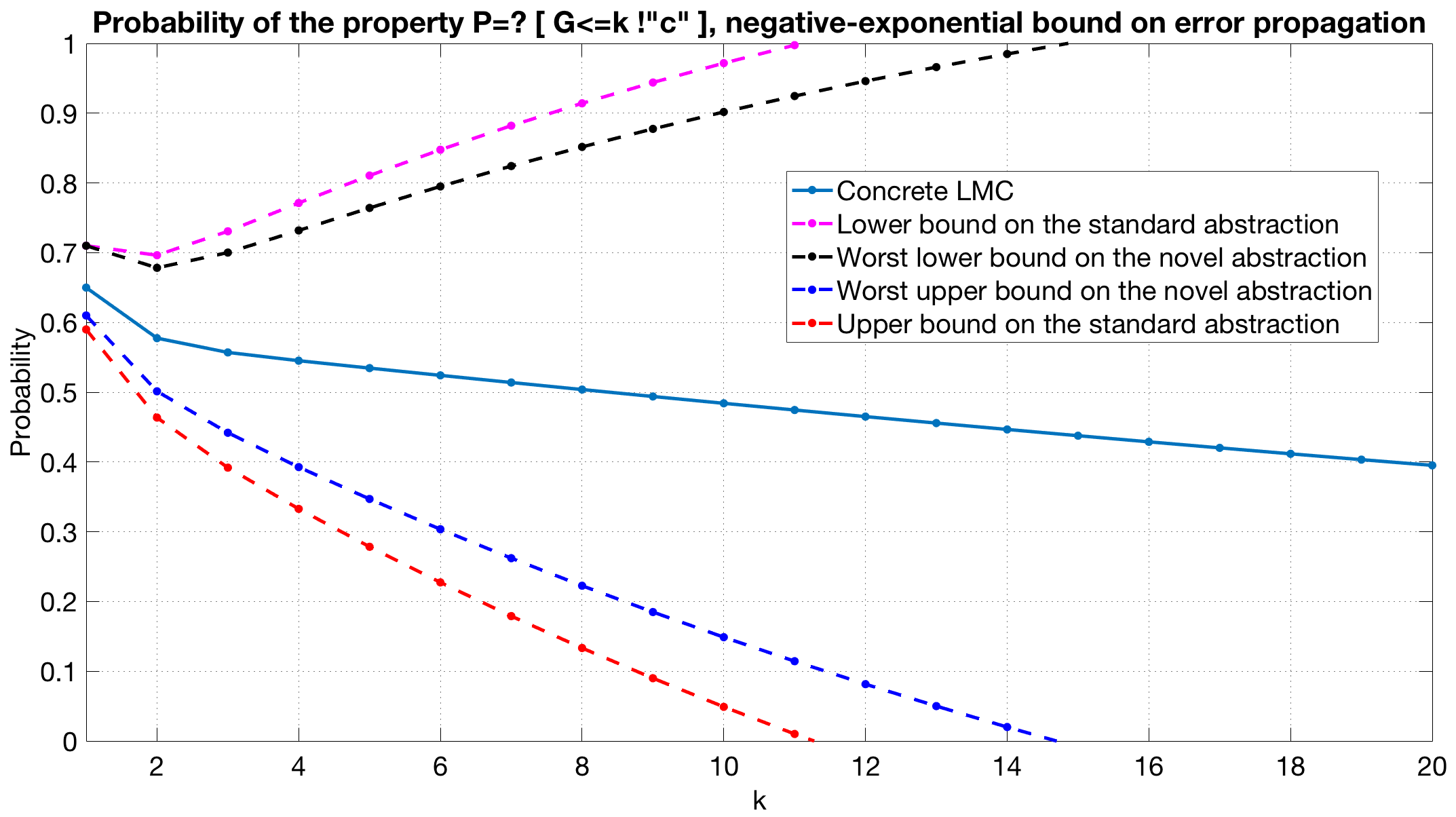}
\caption{Probability of the property $\mathbb{P}_{=?}[\textbf{G}^{\leq k}\neg c]$, with the negative-exponential bound on error propagation}\label{fig:prism}
\end{figure}

So, Fig.~\ref{fig:prism} 
shows the results of querying the probability of $\textbf{G}^{\leq k}\neg c$ 
for the different models in PRISM, where the propagation error is derived as in 
the end of Sect.~\ref{sec:abstractions_novel}, and the semantics for the MDPA are 
the same as in IMDPA (cf. Definition~\ref{def:imdpa_semantics}). It is clear that 
even for the worst optimal error of the MDPA corresponding to $\beta_a\!=\!0.5$, 
the novel abstraction presents a considerable improvement over the standard approach 
to abstractions.

\vspace*{1mm}
\balance                                
\appendix
\section{Proof of Lemma~\ref{l1}}\label{appendix:lemma1}
We proceed by contradiction. 

Let $r^i_{\!j}$ be a vector corresponding to a selection of element $s^i_{\!j}$.
Suppose that its associated error $\varepsilon^i_{\!j} \!<\! \frac{1}{2}\varepsilon_{\max}^i$. 
Let $a, b$ be such that $\infnorm{r_{\!a}^i\!-\!r_{\!b}^i} \!=\! \varepsilon_{\max}^i$. 
Then $\infnorm{r_{\!j}^i\!-\!r_{\!a}^i} \!\leq\! \varepsilon^i_{\!j} \!<\! \frac{1}{2}\varepsilon_{\max}^i$, and
$\infnorm{r_{\!j}^i\!-\!r_{\!b}^i} \!\leq\! \varepsilon^i_{\!j} \!<\! \frac{1}{2}\varepsilon_{\max}^i$. But 
we have that
$$\infnorm{r_{\!a}^i\!-\!r_{\!b}^i} \!=\! \infnorm{(r_{\!a}^i \!-\! r_{\!j}^i) \!+\! (r_{\!j}^i\!-\!r_{\!b}^i)} \!\leq\! 
\infnorm{r_{\!j}^i\!-\!r_{\!a}^i} + \infnorm{r_{\!j}^i\!-\!r_{\!b}^i},$$ 
which is less than 
$\frac{1}{2}\varepsilon_{\max}^i \!+\! \frac{1}{2}\varepsilon_{\max}^i \!=\! \varepsilon_{\max}^i$. 

This is a contradiction, 
since $\infnorm{r_{\!a}^i\!-\!r_{\!b}^i} \!=\! \varepsilon_{\max}^i$. 

So, it follows that $\varepsilon_{\!j}^i \!\geq\! \frac{1}{2}\varepsilon_{\max}^i$. \qed
\section{Proof of Proposition~\ref{prop2}}\label{appendix:prop2}
If $[u^i,v^i]_{\mathrm{opt}}$ is empty, then 
either $\sum_{k=1}^{m} (v^i_k \!-\! \beta_i) \!>\! 1$, or $\sum_{k=1}^{m} (u^i_k + \beta_i) \!<\! 1$. 
Suppose first that $\sum_{k=1}^{m} (v^i_k \!-\! \beta_i) \!>\! 1$. 
Then $r^i$ as in \eqref{eq:optimal_transition_set} 
is non-stochastic, $\max\limits_{1 \leq j \leq n_i}\infnorm{r^i\!-\!r^i_{\!j}} \leq \beta_i$. 
Let $\gamma_i^{\prime}\!=\!
\frac{\sum_{k=1}^{m} (v_k^i\!-\!\beta_i)-1}{m}$, then
$r_{\ast}^i \!=\! (v_1^i \!-\! \beta_i \!-\!  \gamma_i^{\prime}, \dots, v_{m}^i \!-\! \beta_i \!-\! \gamma_i^{\prime}) $ is stochastic,
$\varepsilon_{\ast}^i\!=\!\max\limits_{1\leq j \leq n_i}\infnorm{r_{\ast}^i\!-\!r^i_{\!j}} \!\leq\!\beta_i \!+\! \gamma_i^{\prime} \!=\!
\frac{\sum_{k=1}^{m} v_k^i\!-\!1}{m}$.
Moreover, if 
$\sum_{k=1}^{m} (v^i_k \!-\! \beta_i) \!>\! 1$, then also 
$\sum_{k=1}^{m} (u^i_k \!+\! \beta_i) \!>\! 1$.

Rearrangement of these inequalities brings 
$\frac{1 - \sum_{k=1}^{m} u_k^i}{m} \!<\! \beta_i \!<\! \frac{\sum_{k=1}^{m} v_k^i-1}{m}$,
and
$\varepsilon_{\ast}^i \!\leq\! \max \left\{\frac{1 - \sum_{k=1}^{m} u_k^i}{m}, \frac{\sum_{k=1}^{m} v_k^i-1}{m}\right\}$.

\vspace*{-1mm}
Now, suppose instead that $\sum_{k=1}^{m} (u^i_k + \beta_i) \!<\! 1$. 
Then let $\gamma_i^{\prime} \!=\! 
\frac{1 \!-\! \sum_{k=1}^{m} (u_k^i\!+\!\beta_i)}{m}$.
In this case $r_{\ast}^i \!=\! (u_1^i \!+\! \beta_i \!+\! \gamma_i^{\prime} , \dots , u_{m}^i \!+\! \beta_i \!+\! \gamma_i^{\prime})$ is stochastic and 
$\varepsilon_{\ast}^i\!=\!\max\limits_{1\leq j \leq n_i}\infnorm{r_{\ast}^i\!-\!r^i_{\!j}} \!\leq\!\beta_i \!+\! \gamma_i^{\prime} \!=\!$
$\frac{1 - \sum_{k=1}^{m} u_k^i}{m}$.
Similarly to above, if $\sum_{k=1}^{m} (u^i_k + \beta_i) \!<\! 1$, then 
$\sum_{k=1}^{m} (v^i_k \!-\! \beta_i) \!<\! 1$
and rearranging gives: 
$\frac{\sum_{k=1}^{m} v_k^i-1}{m} \!<\! \beta_i \!<\! \frac{1 - \sum_{k=1}^{m} u_k^i}{m}$,
and hence: 
$\varepsilon_{\ast}^i \!\leq\! \max \left\{\frac{1 - \sum_{k=1}^{m} u_k^i}{m}, \frac{\sum_{k=1}^{m} v_k^i-1}{m}\right\}$. \qed

\section{Proof of Lemma~\ref{size1}}\label{appendix:lemma2}
For the backwards direction, suppose that $\sum_{k=1}^{m} t_k \!=\! 1$. Then 
$t\!=\!(t_1, \dots , t_{m}) \!\in\! T$. 
If $a\!=\!(a_1,\dots , a_{m}) \!\in\! T$, where $a \!\neq\! t$, then for all $1 \!\leq\! j \!\leq\! {m}$, 
$a_j \!\geq\! t_j$, and for some 
$1 \!\leq\! k \!\leq\! m$, $a_k\!>\!t_k$. This implies that $\sum_{k=1}^{m} a_k \!>\! 1$, and hence $a \!\notin\! T$, a contradiction. 
Thus $a\!=\!t$, and hence $T\!=\!\{t\}$. So $\vert T \vert \!=\!1$. The argument for $\sum_{k=1}^{m} w_k \!=\!1$ is similar. 
Now, suppose that $\sum_{k=1}^{m} t_k \!<\! 1 \!<\! \sum_{k=1}^{m} w_k$, and $T$ has only one free element. 
Without loss of generality, assume that the first element of $T$ is free, so $t_1 \!<\! w_1$, 
and for all $1 \!<\! j \!\leq\! m$, $t_j\!=\!w_j$. Then 
$t_1 \!<\! 1 \!-\!\! \sum_{j\neq 1} t_j \!=\! 1 \!-\! \sum_{j\neq 1} w_j \!<\! w_1$, and hence 
$(1\!-\!\!\sum_{j\neq 1} t_j, t_2, \dots , t_{m}) \!\in\! T$. Clearly, as the first is the only free element of $T$, 
$T \!=\! \{(1\!-\!\!\sum_{j\neq 1} t_j, t_2, \dots , t_{m} )\}$.
For the forwards direction, we go by contraposition, assuming that none of the conditions hold. 
If either $\sum_{k=1}^{m} t_k \!>\! 1$ or $\sum_{k=1}^{m} w_k \!<\! 1$, then $T$ must be empty. 
So we must suppose that $\sum_{k=1}^{m} t_k \!<\! 1 \!<\! \sum_{k=1}^{m} w_k$ and that
$T$ has $\ell\!\geq\!2$ free elements. 
Without loss of generality, formally we write that 
$T \!=\! ([t_1,w_1], \dots , [t_{\ell},w_{\ell}], t_{\ell+1},\dots, t_{m})$. 
Then, for any $a\!=\!(a_1,\dots, a_{m}) \!\in\! T$ there is at least one 
$1 \!\leq\! j \!\leq\! \ell$ such that $a_j \!>\! t_j$, and at least one 
$1 \!\leq\! k \!\leq\! \ell$ such that $a_k \!>\! w_k$. So suppose without loss of generality that 
$a_1 \!>\! t_1$ and $a_2 \!<\! w_2$. Let $\varepsilon \!=\! \min \{a_1\!-\!t_1, w_2\!-\!a_2\}$. 
Then also $a^{\prime} \!=\! (a_1 \!-\! \varepsilon , a_2 \!+\!\varepsilon, a_3, \dots , a_{m}) \!\in\! T$, 
and hence $\vert T \vert \!>\! 1$. \qed  
\bibliography{bibliography}             
\end{document}